\documentclass[a4paper,11pt]{article}
\pdfoutput=1 % if your are submitting a pdflatex (i.e. if you have
\usepackage{jinstpub} % for details on the use of the package, please
\usepackage{epsfig}
\usepackage{epstopdf}
\usepackage{graphicx}
\usepackage{subfigure}
\usepackage{hyperref}
\usepackage{caption}
\usepackage[section,above,below]{placeins}
\usepackage{multirow}
\usepackage{hhline}% http://ctan.org/pkg/hhline
\usepackage{makecell}

\title{\boldmath Characterization of a $^{109}$Cd gamma-ray source for the two-phase argon detector}

\author[a,b]{A.~Bondar,}
\author[a,b]{A.~Buzulutskov,}
\author[b]{A.~Dolgov,}
\author[a]{A.~Legkodymov,}
\author[a,b]{V.~Nosov,}
\author[a,b,1]{V.~Oleynikov,\note{Corresponding author.}}
\author[a,b]{V.~Porosev,}
\author[a,b]{E.~Shemyakina,}
\author[a,b]{A.~Sokolov}

\affiliation[a]{Budker Institute of Nuclear Physics, \\Lavrentiev ave. 11, Novosibirsk 630090, Russia}
\affiliation[b]{Novosibirsk State University,\\ Pirogova st. 2, Novosibirsk 630090, Russia}

\emailAdd{V.P.Oleynikov@inp.nsk.su}

\abstract{
At present, a two-phase argon detector is being developed in our laboratory for dark matter search and low-energy neutrino experiments.
To calibrate its energy scale a $^{109}$Cd gamma-ray source was used.
In this work a detailed emission spectrum of the $^{109}$Cd source was measured using YAP:Ce scintillator and high-purity germanium (HPGe) detectors.
It is shown that the $^{109}$Cd source, equipped with a W substrate and a Cu filter, can provide a complete set of gamma-ray lines, ranging from 8 to 90 keV, for the energy calibration of two-phase detectors.
These measurements allowed us to successfully quantify the shape of the amplitude spectra observed in the two-phase argon detector when irradiated with the $^{109}$Cd source.
}


\begin{document}
\maketitle
\flushbottom
\newcommand*{\doi}[1]{\href{http://dx.doi.org/#1}{doi: #1}}

\section{Introduction}
%\hspace*{\parindent}
The main purpose of this work is to provide reference data on the emission spectrum of a $^{109}$Cd gamma-ray source.
Such a radioactive source is used to calibrate the energy scale of a two-phase argon detector, which is being developed in our laboratory for dark matter search and low-energy neutrino experiments \cite{Buzulutskov2012,XRayYield16,PhotonEmission2017,IonYield17,CRADELGap17,FurtherStudiesEL2017,RevealingNBr2018}.

This study was motivated by the recent observations in the two-phase argon detector \cite{CRADELGap17}: the standard gamma-ray lines expected from a $^{109}$Cd source, at 22-25 and 88 keV \cite{Heath1964,Rittersdorf2007,PDG2017CommonlyUsedRadioactiveSources,Cd109isotopeScheme,NucleardataNuclearLuSe}, were accompanied by additional lines around 60 keV.
In this work, we carefully measured the emission spectrum of the $^{109}$Cd source using YAP:Ce scintillator and high-purity germanium (HPGe) detectors.
This allowed us to successfully quantify the $^{109}$Cd emission spectrum measured in the two-phase argon detector.
To the best of our knowledge, this is the first time that such a detailed $^{109}$Cd emission spectrum has been published in the literature.

%\clearpage
%\FloatBarrier
\section{Description of $^{109}$Cd radioactive source}
\begin{figure}[h!]
	\center{\includegraphics[width=0.7\textwidth]{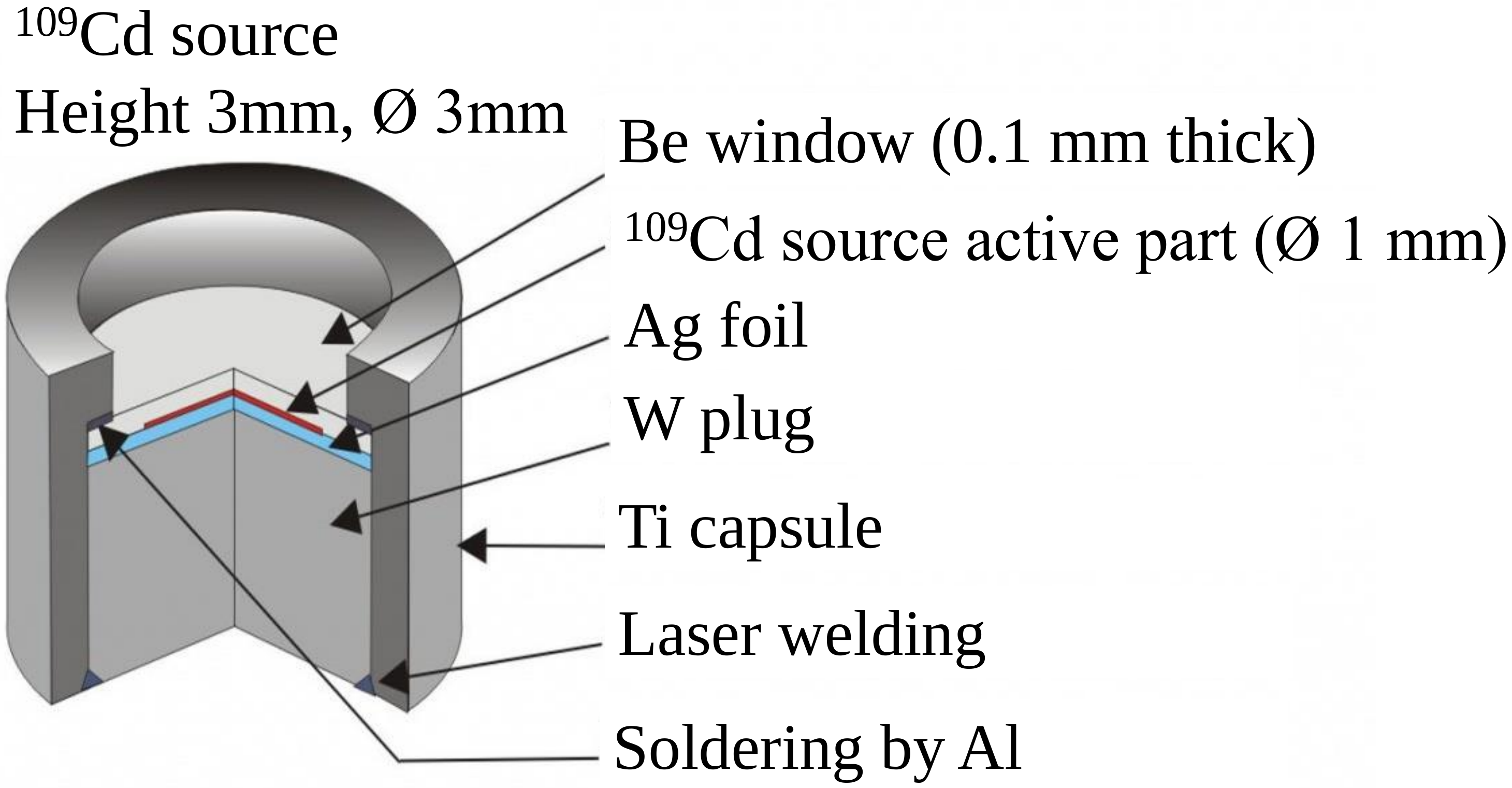}}
	\caption{3D view of the $^{109}$Cd source.}
	\label{fig_Cd109_isotop_box}
\end{figure}

A $^{109}$Cd isotope decays into $^{109}$Ag with a half-life of 461 days, emitting a number of gamma-ray lines in the energy range of 20-90 keV \cite{Heath1964,Rittersdorf2007,PDG2017CommonlyUsedRadioactiveSources,Cd109isotopeScheme,NucleardataNuclearLuSe}.
A schematic view of the $^{109}$Cd source used in our measurements is shown in figure~\ref{fig_Cd109_isotop_box}.
The source was produced by Cyclotron company and had an activity of 2.5$\cdot$10$^8$ Bq in April 2014 \cite{Cd109isotope}.

The radionuclide $^{109}$Cd was electroplated on a Ag foil.
The foil with the radionuclide was hermetically sealed in a Ti capsule.
The hermetization of the capsule was provided by laser welding.
From one side of the radionuclide a W plug (substrate) was mounted to absorb the radiation; and on the other side there was a thin Be window.

%\FloatBarrier
%\subsection{HPGe detector}
\section{Measurements with HPGe detector}
The most precise emission spectrum of the $^{109}$Cd source was measured using a HPGe detector \cite{Knoll2000}, namely that of GUL/EGX 10-05 Canberra \cite{CanberraHPGe}.
The measurement conditions were similar to that of the two-phase detector in \cite{CRADELGap17}.
In particular, Al and acrylic filters, 3 mm thick each, were placed between the $^{109}$Cd source and the detector.

\begin{figure}[ht!]
	\centering
	{
	\begin{minipage}[h]{0.79\linewidth}\centering
		\includegraphics[width=1\linewidth]{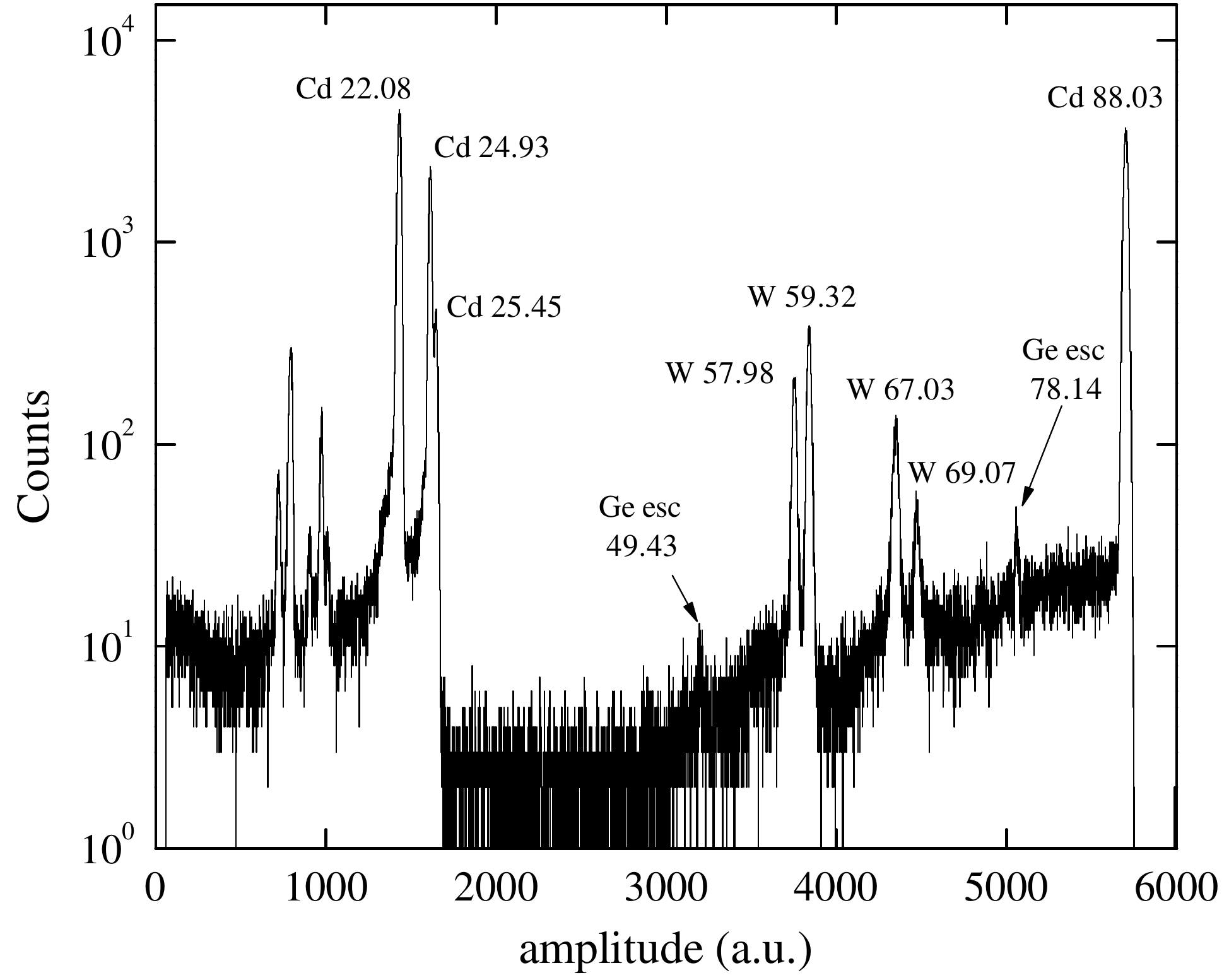} 
	\end{minipage}
	\vfill
	\vspace{4ex}
	\begin{minipage}[h]{0.79\linewidth}\centering
		\includegraphics[width=1\linewidth]{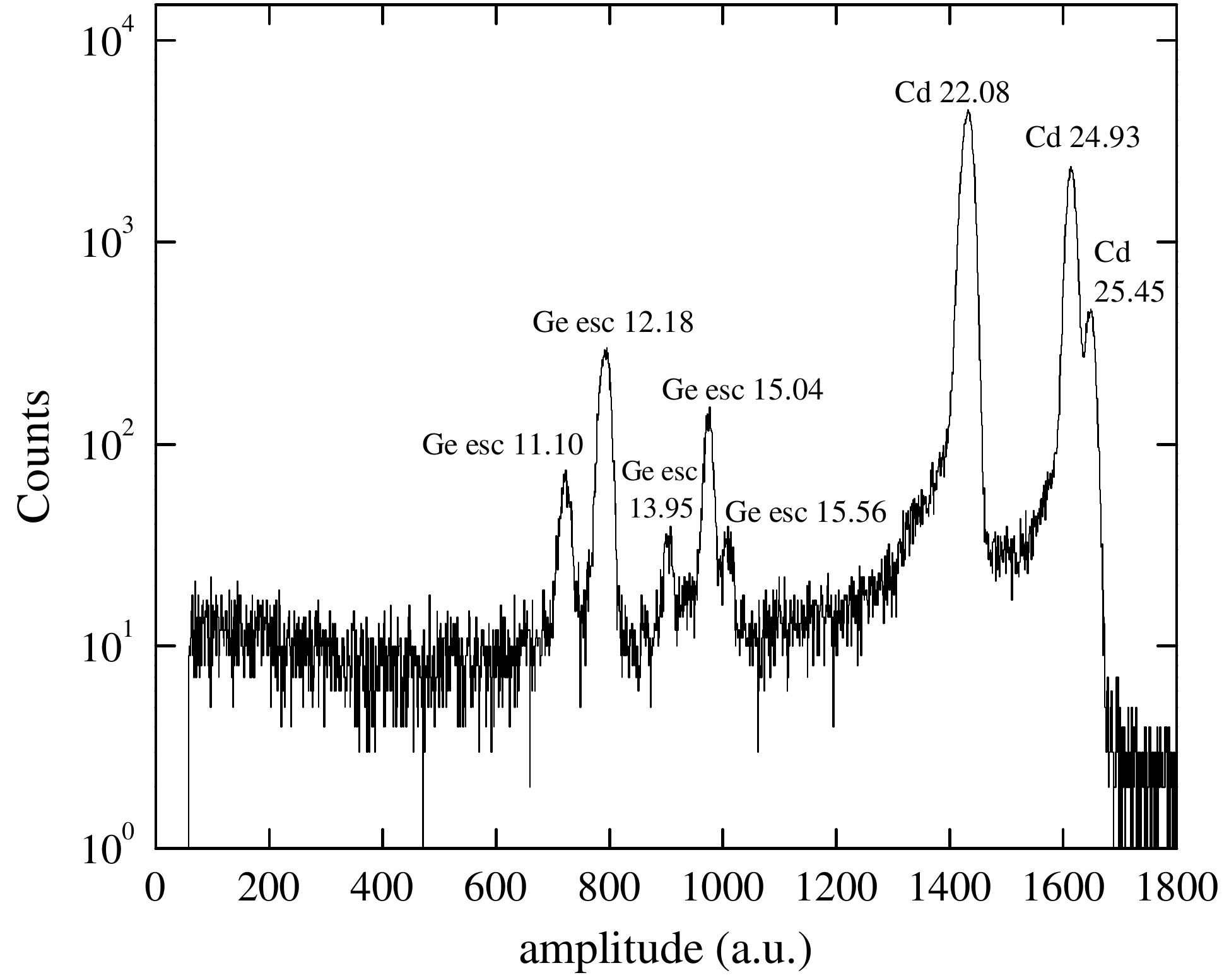}
	\end{minipage}
	}
	\caption{$^{109}$Cd emission spectrum after Al and acrylic filter (3 mm thick each) measured using a HPGe detector: the whole spectrum (top) and its low energy part (bottom). The X-ray characteristic lines, at 60-70 keV, are due to the W substrate of the $^{109}$Cd source.}
	\label{171226_Cd109_spectrum_from_Ge_det}  
\end{figure}

Figure~\ref{171226_Cd109_spectrum_from_Ge_det} shows the resulting amplitude spectrum.
Energy, origin, intensity and energy resolution for identified peaks (except escape peaks) are presented in table~\ref{tab.Cd109_specrtum}.

\begin{table}[h]
\captionof{table}{Identified peaks in the $^{109}$Cd emission spectrum (except escape peaks) measured using the HPGe detector.
The intensity and the energy resolution were calculated by Gaussian approximation, where the background continuum was subtracted.
The peak energies were taken from \cite{Cd109isotopeScheme,X_Ray_Data_Booklet}.}\label{tab.Cd109_specrtum}
\begin{center}
\begin{tabular}{|c|c|c|c|}
\hline
Energy (keV) & Origin &  \makecell{ Area, normalized \\ to 88.03~keV line} & $\sigma / E \cdot 10^{3}$   \\
\hline
88.03 & $^{109}$Cd $\rightarrow$ $^{109}$Ag & {1} & 2.2 \\
%\cline{2-4}
69.07 & W k$\beta_2$ & {0.01} & 2.8 \\
%\cline{2-4}
67.03 & W k$\beta_1$ + k$\beta_3$ & {0.037} & 3.6 \\
%\cline{2-4}
59.32 & W k$\alpha_1$ & {0.092} &  2.9 \\
%\cline{2-4}
57.98 & W k$\alpha_2$ & {0.049} & 2.9 \\
%\hline
25.45 & Ag k$\beta_2$ & 0.09 & 5.1 \\
%\cline{2-4}
24.93 & Ag k$\beta_1$ + k$\beta_3$ & 0.44 & 5.3 \\
%\cline{2-4}
22.08 & Ag k$\alpha_1$ + k$\alpha_2$ & 0.97 & 7.0 \\
\hline
\end{tabular}
%\vspace*{2em}
\end{center}
\end{table}

First of all, the primary peaks due to $^{109}$Cd and $^{109}$Ag were identified at 88.03, 25.45, 24.93 and 22.08 keV \cite{Cd109isotopeScheme}.
In the range of 60-70 keV four secondary peaks due to X-ray fluorescence (characteristic lines) of W were identified at 69.07, 67.03, 59.32 and 57.98 keV.
All other identified peaks are due to X-rays escaped from the HPGe detector with the escaped energies of 9.87(Ge k$\alpha_1$ + Ge k$\alpha_2$) and 10.98(Ge k$\beta_1$ + Ge k$\beta_3$) keV, that is, these are the escape peaks.
For example, the 11.10 keV peak is the escape peak induced by absorption of a primary 22.08 keV photon, where a secondary (fluorescent) 10.98 keV photon escaped the detector.

In addition to the peaks, there were two regions with continuous spectra.
The first continuum, in the range of 25-88 keV, is produced when the 88.03 keV photons or W fluorescence photons scatter on the surrounding materials and then come back to the detector.
The second continuum, in the range of 0-22 keV, is produced when the gamma-rays scatter inside the detector, leaving there some energy, and then escape the detector.

%\afterpage{\FloatBarrier}
%\FloatBarrier
%\clearpage
%\subsection{YAP:Ce scintillator coupled with PMT}
\section{Measurements with YAP:Ce scintillator detector}
In additional to the measurements with the HPGe detector, the $^{109}$Cd emission spectrum was measured using a YAP:Ce (YAlO$_3$, doped with Ce) scintillator coupled to a photomultiplier tube (PMT) Hamamatsu R10233.
To match the emission spectra of the scintillator with the quantum efficiency spectrum of the PMT, the crystal was coated with a
POPOP + PPO-based wavelength shifter (WLS) \cite{Babichev2015}.
The $^{109}$Cd source was separated from the scintillator with an attenuation filter made from a Cu foil (0.17 mm thick).

\begin{figure}[ht!]
	\center{\includegraphics[width=0.7\textwidth]{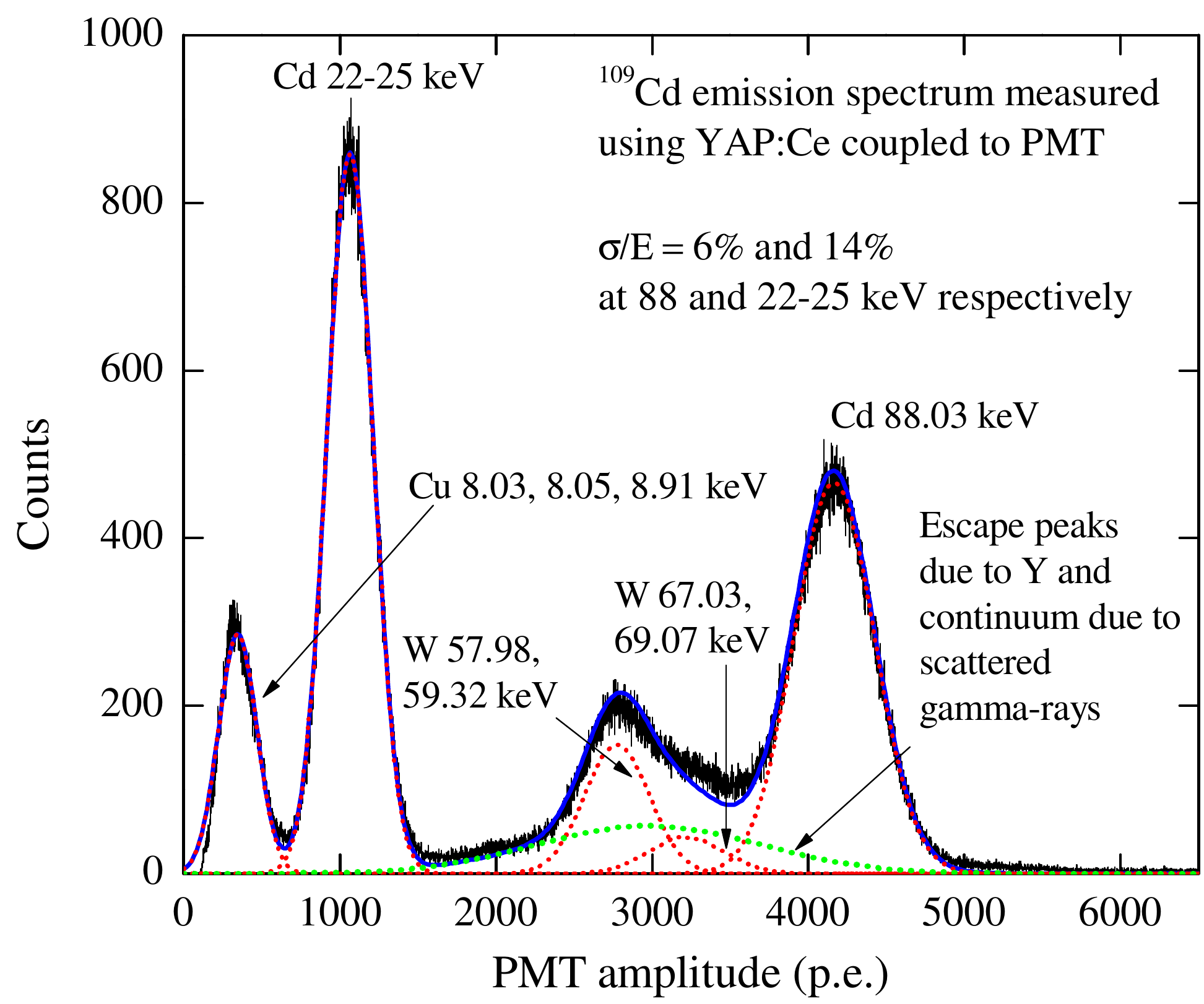}}
	\caption{$^{109}$Cd emission spectrum measured using a YAP:Ce detector. The $^{109}$Cd source was equipped with a Cu attenuation filter.}
	\label{fig_YAPCe_PMT_spectrum}
\end{figure}

Figure~\ref{fig_YAPCe_PMT_spectrum} shows the measured amplitude spectrum.
The contributions of different gamma-ray and X-ray lines are shown by the red and green dotted lines, while the blue solid line is the overall fit of the experimental data.
On the one hand, the amplitude spectrum is similar to that of HPGe detector: 
the $^{109}$Cd gamma-lines and the X-ray characteristic lines of W are well distinguished.
On the other hand, a Cu characteristic line at 8 keV has emerged, while those of the escape peaks due to Ge have disappeared.
In addition, the escape lines due to Y and the continuum due to scattered radiation were observed: their total contribution is shown in green dotted line.
Since the energy resolution of YAP:Ce detector is worse compared to that of HPGe detector, the softer Cd lines have merged, as well as those of W.

%\subsection{Two-phase argon detector}
\section{Measurements with two-phase detector}

\begin{figure}[ht]
	\center{\includegraphics[width=0.7\textwidth]{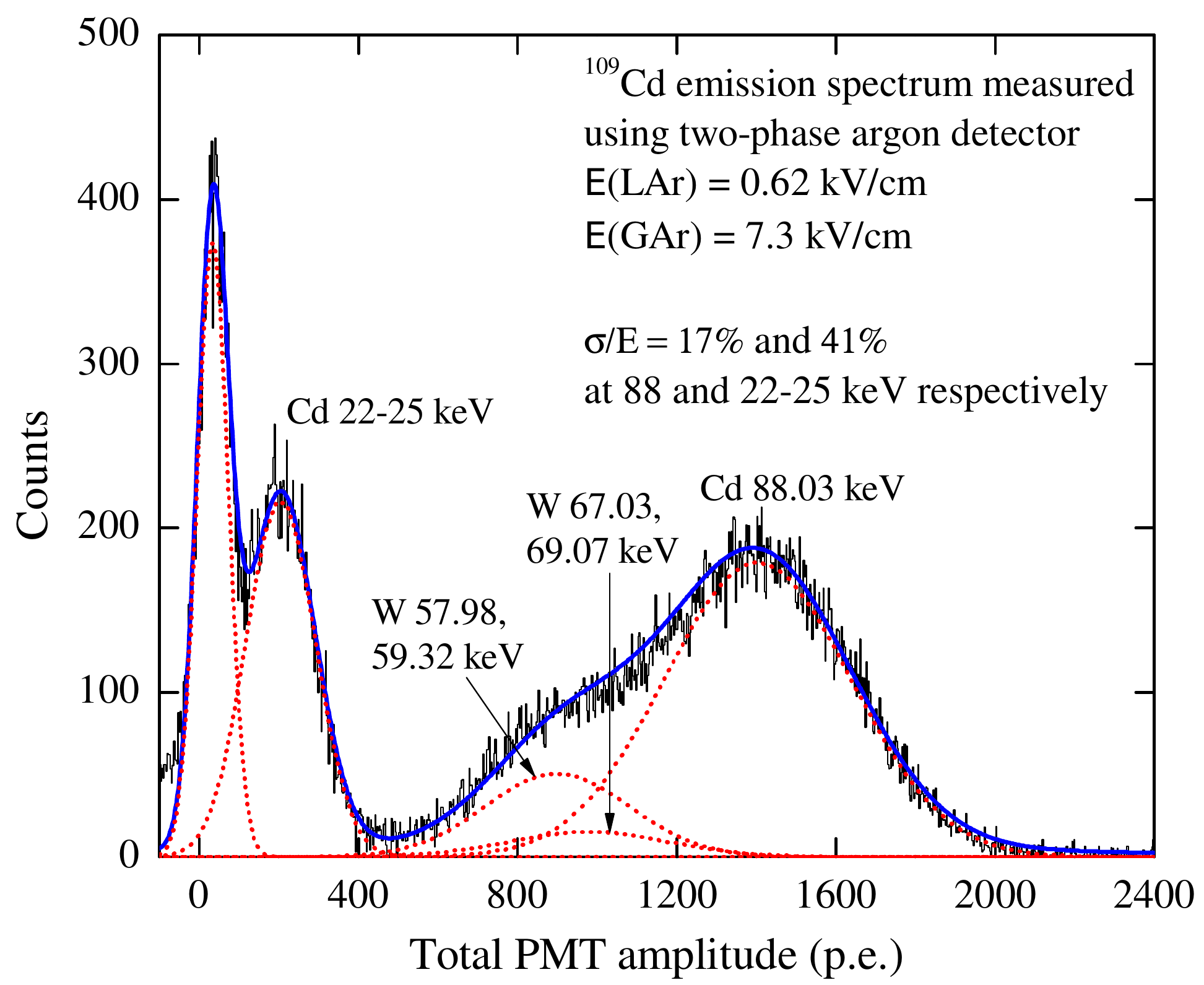}}
	\caption{$^{109}$Cd emission spectrum measured using a two-phase argon detector \cite{CRADELGap17,FurtherStudiesEL2017}.}
	\label{fig_CRAD_Cd109}
\end{figure}

Figure~\ref{fig_CRAD_Cd109} shows the $^{109}$Cd emission spectrum measured using a two-phase argon detector with electroluminescence gap described elsewhere \cite{CRADELGap17,FurtherStudiesEL2017}.
The energy resolution of the two-phase detector was somewhat worse than that of YAP:Ce.
Accordingly, the spectrum shape in the two-phase detector was generally similar to that of YAP:Ce, despite the worst separation of the W and Cd lines.
The overall spectrum is well described by the model, including the Cd lines, at 22-25 and 88 keV, and W characteristic lines, at 58-60 and 67-69 keV. 
It should be remarked that in \cite{CRADELGap17} the W-lines in the amplitude spectrum were incorrectly interpreted as those being due to $^{241}$Am source.
Thus in the current work we have corrected that misinterpretation.

% has the same effect as \newpage but restricts floats as well
%\clearpage
\section{Conclusions}

In this work the emission spectrum of a $^{109}$Cd source was characterized using YAP:Ce scintillator and high-purity germanium (HPGe) detectors.
These measurements allowed us to successfully quantify the shape of the amplitude spectrum measured in a two-phase argon detector when irradiated with the $^{109}$Cd source.
Along with the main gamma-ray lines of $^{109}$Cd source, at 22-25 and 88 keV, the associated characteristic lines of W (due to W substrate), at 60-70 keV, and those of Cu (due to Cu filter), at 8 keV, were observed.
Thus, the $^{109}$Cd source, equipped with a W substrate and a Cu filter, can provide a complete set of gamma-ray lines, ranging from 8 to 90 keV, for the energy calibration of two-phase detectors.
To the best of our knowledge, this is the first time that such a detailed Cd emission spectrum has been presented in the literature.

\acknowledgments
This study was supported by Russian Science Foundation (project no. 16-12-10037).
The part of the work relating to the measurement of the $^{109}$Cd emission spectrum using the HPGe detector was done using the infrastructure of the Shared-Use Center "Siberian Synchrotron and Terahertz Radiation Center (SSTRC)" based on VEPP-4M of BINP SB RAS.

\bibliographystyle{JHEP}
\bibliography{mybibliography}

\end{document}